\documentclass[aps,prd,epsf,showpacs,amsmath,amssymb,graphics,preprint]{revtex4}
\usepackage{graphicx}
\usepackage{dcolumn}
\usepackage{bm}
\usepackage{color}

\newcommand{\be}{\begin{equation}}
\newcommand{\ee}{\end{equation}}
\newcommand{\ba}{\begin{eqnarray}}
\newcommand{\ea}{\end{eqnarray}}
\newcommand{\ban}{\begin{eqnarray*}}
\newcommand{\ean}{\end{eqnarray*}}

\newcommand{\A}{\ensuremath{{\cal A}}}
\newcommand{\B}{\ensuremath{{\cal B}}}
\newcommand{\C}{\ensuremath{{\cal C}}}

\begin{document}

\title{How small can an over-spinning body be in general relativity?}
\author{$^1$Ken-ichi Nakao\footnote{Electronic address: knakao@sci.osaka-cu.ac.jp}
$^2$Masashi Kimura\footnote{Electronic address: M.Kimura@damtp.cam.ac.uk}, 
$^3$Tomohiro Harada\footnote{Electronic address: harada@rikkyo.ac.jp}, 
$^4$Mandar Patil\footnote{Electronic address: mandar@iucaa.ernet.in}, 
$^5$Pankaj S. Joshi\footnote{Electronic address: psj@tifr.res.in}
}

\affiliation{$^1$Department of Mathematics and Physics, Graduate School of Science, Osaka City University, Osaka 558-8585, Japan. \\ 
$^2$DAMTP, Centre for Mathematical Sciences, University of Cambridge,
Wilberforce Road, Cambridge CB3 0WA, United Kingdom. \\
$^3$Department of Physics, Rikkyo University, Toshima-ku, Tokyo 171-8501 Japan. \\
$^4$Inter University Center for Astronomy and Astrophysics Post Bag 4, Ganeshkhind, Pune-411007, India. \\
$^5$Tata Institute of Fundamental Research, Homi Bhabha Road, Mumbai 400005, India.
}

\begin{abstract}
The angular momentum of the Kerr singularity should not be 
larger than a threshold value so that it is enclosed by an event horizon: 
The Kerr singularity with the angular momentum exceeding 
the threshold value is naked. 
This fact suggests that if the cosmic censorship exists in our Universe, 
an over-spinning body without releasing its angular momentum cannot collapse  
to spacetime singularities. 
A simple kinematical estimate of two particles approaching each other 
supports this expectation and  
suggests the existence of a minimum size of an over-spinning body. 
But this does not imply that the geometry near the naked singularity 
cannot appear. By analyzing initial data, i.e., a snapshot of a 
spinning body, we see that an over-spinning body 
may produce a geometry close to the Kerr naked singularity 
around itself at least as a transient configuration. 

\end{abstract}
\preprint{OCU-PHYS-402}
\preprint{AP-GR-110}

\pacs{04.20.-q, 04.20.Cv, 04.20.Dw, 04.20.Ex, 04.25.D-}

\maketitle

\section{Introduction}

It is a well-known fact that the Kerr singularity of mass $M$ is enclosed 
by an event horizon if and only if 
its angular momentum $J$ is not larger than a threshold value $J_{\rm max}:=GM^2/c$, 
where $G$ and $c$ are Newton's gravitational constant and the speed of light, 
respectively: the Kerr singularity with $J>J_{\rm max}$ is necessarily naked 
(see e.g., Ref.\cite{Wald}). If the cosmic censorship conjecture which states 
that the spacetime singularity produced by the physically reasonable gravitational 
collapse is enclosed by the event horizon\cite{penrose1969,penrose1979} is true, 
an over-spinning body cannot collapse to spacetime 
singularities if it does not release its angular momentum. 
A simple kinematical estimate supports this expectation: If we impose a 
condition on the total angular momentum $J>J_{\max}$, 
the impact parameter $b$ of two test particles without 
any interaction in Minkowski spacetime is bounded below as $b>2GE/c^4$, 
where $E$ is the total energy of the system. 
However, it is a very nontrivial question whether an over-spinning body can be so 
small even for a moment that the geometry around it is 
almost equal to that of the domain very near the naked singularity in the over-spinning Kerr spacetime. 

There are several studies of the gravitational collapse of an over-spinning 
body\cite{Nakamura,Nakamura-Sato,Stark-Piran,Abrahams_etal,Duez_etal,Giacomazzo_etal}.   
Their results imply that the over-spinning body does not form spacetime singularities 
without releasing its angular momentum. Here it should be noted that all 
of these studies focus on the situations in which the over-spinning body is 
gravitationally bound or at most marginally bound initially. 
Although the gravitationally bound 
initial condition is a reasonable assumption in astrophysical studies, 
it is too restrictive to get an insight into this fundamental question in general relativity. 
It is also necessary to consider more general situations, e.g.,  
the kinetic energy dominant implosion.  

There are several studies of the systems with kinetic energy dominant initial 
conditions. Their purpose is not to resolve  
the astrophysical problems but to understand 
the black hole formation through the high energy collision of elementary particles, 
however these studies have not paid attention to  
the present issue\cite{Washik,Shibata,HE-BH-1,HE-BH-2}.

From the point of view of the cosmic censorship, 
Wald studied the motion of a test particle around an extreme Kerr black hole 
and showed that if the sum of the angular momentum of the test particle and 
the Kerr black hole exceed the threshold value $J_{\rm max}$, the particle cannot 
enter the black hole\cite{wale-paper}. This is the case including gravity 
that suggests the existence of a lower bound on the size of over-spinning body. 
Later, Jacobson and Sotirious showed that if the angular momentum of the 
black hole is a bit less than the threshold value, a test particle can enter the 
black hole even though the sum of the angular momenta of the black hole and 
the particle exceeds the threshold\cite{Jacobson}. However, the study by Barausse, Cardoso and  Khanna 
suggests that if the self-force of the particle is taken into account, the particle cannot enter 
the black hole if the total angular momentum exceeds the threshold\cite{Barausse1,Barausse2}.
These studies seem to imply that there is a lower bound on the size of an over-spinning body. 
But, these results may merely imply the stability of the horizon. 
As far as we know, there is no study of the situations with no horizon. 

In order to get an answer to this question, 
we do not need to investigate dynamical processes but it is sufficient to only study 
the initial data of the Cauchy problem in general relativity. In this paper, we set up 
the initial data of an axisymmetric infinitesimally thin shell with the topology of ${\bf S}^2$ by numerically 
solving the constraint equations in the Einstein equations. We assume that the outside of the shell 
is identical to a spacelike hypersurface of the Kerr 
spacetime; such initial data was discussed by Corvino and Schoen\cite{CS}.  
We assume that the inside of the shell is vacuum regular space.     

The shell is assumed to be located at the constant radial coordinate $r=R$ of the Boyer-Lindquist coordinates 
which cover the Kerr domain outside the shell. We investigate how small $R$ can be in the case 
of the over-spinning shell, $J>J_{\rm max}$, under the weak, strong and dominant energy 
conditions which seem to be reasonable for macroscopic matter fields. 

Hereafter, we adopt the geometrized units $G=c=1$. 
In this paper, the Greek indices represent spacetime 
components, whereas the Latin indices donate the spatial components. 

\section{Constraint equations}

A set of the intrinsic metric $\gamma_{ij}$, the extrinsic curvature $K_{ij}$  
of a spacelike hypersurface $\Sigma_0$, 
and the energy density and the momentum density of matter fields 
can be the initial data of the Cauchy problem in general relativity (see e.g. \cite{Gourgoulhon}).   
We may regard this set as a snapshot of the system. 

The intrinsic metric $\gamma_{ij}$ determines the intrinsic geometry 
of $\Sigma_0$, whereas the extrinsic curvature $K_{ij}$ determines 
how $\Sigma_0$ is embedded in the spacetime manifold. 
The future directed unit vector normal to $\Sigma_0$ is denoted by $n_\mu=(-\alpha,0,0,0)$, 
where $\alpha$ is called the lapse function.  As usual, we denote the spacetime metric 
by $g_{\mu\nu}$. 
The projection operator to $\Sigma_0$ is defined as 
\begin{equation}
B_\mu^\nu:=\delta_\mu^\nu+n_\mu n^\nu~~~~{\rm or~~equivalently}~~~~B_{\mu\nu}:=g_{\mu\nu}+n_\mu n_\nu,
\end{equation}
and we have $\gamma_{ij}=B_{ij}$.  The extrinsic curvature is defined as
\begin{equation}
K_{\mu\nu}:=-B_\mu^\alpha\nabla_\alpha n_\nu.
\end{equation}
From this definition, we can see that $K_{\mu\nu}$ is the spatial tensor, i.e., 
$K_{\mu\nu}n^\nu=0=n^\mu K_{\mu\nu}$, and is rewritten in the form
\begin{equation}
K_{ij}=-\frac{1}{2\alpha}\left(\frac{\partial\gamma_{ij}}{\partial t}-D_i\beta_j-D_j\beta_i\right),
\end{equation}
where $D_i$ is the covariant derivative with respect to $\gamma_{ij}$, and 
$\beta_i:=g_{0i}$ is called the shift vector. 
The energy density $\rho$ and the momentum density $J^i$ for normal line observers are defined as
\begin{eqnarray}
\rho:=T_{\mu\nu} n^\mu n^\nu, \\
J_i:=-T_{\mu\nu}n^\mu B^\nu_i,
\end{eqnarray}
where $T_{\mu\nu}$ is the stress-energy tensor of matter or radiation fields. 

The initial values must satisfy  the constraint equations which are the 
time-time component and time-space components of the Einstein equations; the 
former is called the Hamiltonian constraint, and the latter the momentum 
constraint. These are written in the form
\begin{eqnarray}
^3 R-K^{ij}K_{ij}+K&=&16\pi \rho, \label{HC}\\
D_j (K_i^j-\delta_i^j K)&=&8\pi J_i, \label{MC}
\end{eqnarray}
where $^3 R$ is the Ricci scalar of $\gamma_{ij}$, and $K:=\gamma^{ij}K_{ij}$.

\section{Initial data: a snapshot of a rapidly rotating shell}

As mentioned, we set up the initial data of a rapidly rotating infinitesimally thin shell 
with the spherical topology ${\bf S}^2$. The energy density and the momentum density confined on the shell 
are not fixed prior to solving the constraint equations (\ref{HC}) and (\ref{MC}) in the prescription 
we adopt. In this section, we show how to obtain the initial data of $\gamma_{ij}$ and $K_{ij}$ by using 
the conformal decomposition\cite{york1,york2,york3}.  

We assume that the system is axisymmetric and its infinitesimal line element is written in the form
\begin{equation}
d\ell^2=\phi^4(r,\theta)\left[\A(r,\theta)dr^2+r^2\B(r,\theta)d\theta^2+r^2\C(r,\theta)\sin^2\theta d\varphi^2\right],
\label{metric}
\end{equation}
where $0\leq r <\infty$, $0\leq\theta\leq \pi$ and $0\leq\varphi <2\pi$ are spherical polar coordinates. 
We assume that the infinitesimally thin shell is located at $r=R=$constant. Hereafter we call 
$$
\lambda_{ij}:={\rm diag}\left[\A,~r^2\B,~r^2\C\sin^2\theta\right]
$$ 
the conformal metric. 

In order to define the ``size" of the shell without any ambiguities, 
we assume that the domain outside the shell is 
exactly the same as the initial data of the over-spinning Kerr spacetime. 
We adopt the Boyer-Lindquist coordinates for the outside Kerr domain, $r>R$, and hence, 
by defining the following three functions,
\begin{eqnarray}
\varDelta(r)&:=&r^2-2Mr+a^2, \\
&& \cr
A(r,\theta)&:=&\left(r^2+a^2\right)^2-a^2\varDelta\sin^2\theta, \\
&& \cr
\varSigma(r,\theta)&:=&r^2+a^2\cos^2\theta,
\end{eqnarray}
the metric functions outside the shell, $r>R$, are given by
\begin{eqnarray}
\phi(r,\theta)&=&1, \\
\A(r,\theta)&=&\frac{\varSigma}{\varDelta}, \\
\B(r,\theta)&=&\frac{\varSigma}{r^2}, \\
\C(r,\theta)&=&\frac{A}{r^2\varSigma}, 
\end{eqnarray}
where $M$ and $a$ are the Arnowitt-Deser-Misner (ADM) mass and the Kerr parameter, respectively. 
Note that the ADM mass corresponds to the total energy, whereas $Ma$ 
is the total angular momentum. 
The nonvanishing components of the extrinsic curvature of $\Sigma_0$ 
in the domain outside the shell, $r>R$, are given by
\begin{equation}
K^{r\varphi}=\frac{1}{2\alpha \A}\frac{\partial\beta^\varphi}{\partial r}~~~~~~{\rm and}~~~~~~
K^{\theta\varphi}=\frac{1}{2r^2 \alpha \B}\frac{\partial\beta^\varphi}{\partial\theta},
\label{exc-out}
\end{equation}
where
\begin{eqnarray}
\alpha(r,\theta)&=&\sqrt{\frac{\varSigma\varDelta}{A}}, \\
&&\cr
\beta^\varphi(r,\theta)&=&-\frac{2aMr}{A}.
\end{eqnarray}
Of course, the above intrinsic metric and the extrinsic curvature satisfy the constraint equations 
(\ref{HC}) and (\ref{MC}) with $\rho=0=J_i$. 
As is well known, there is a ring singularity at $(r,\theta)=(0,\pi/2)$ 
in the Kerr spacetime with Boyer-Lindquist coordinates. 
Hereafter we assume $R>0$ so that no spacetime singularity exists outside the shell. 

We assume that the inside of the shell, $r<R$, is vacuum and regular.  
By defining a smoothed step function as 
\begin{equation}
W(r)=
\left\{
\begin{array}{ll}
0&{\rm for}~0\leq r \leq R-L \\
L^{-36}[(r-R)^6-L^6]^6&{\rm for}~R-L \leq r \leq R \\
1&{\rm for}~R \leq r \\
\end{array}\right.,
\label{W-def}
\end{equation}
with a constant $L$ which satisfies $0<L<R$, we assume that  
the metric functions $\A$, $\B$ and $\C$ are  
\begin{eqnarray}
\A(r,\theta)&=& \frac{\varSigma}{\varDelta}W+\Psi(1-W), \\
\B(r,\theta)&=&\frac{\varSigma}{r^2}W+\Psi(1-W), \\
\C(r,\theta)&=& \frac{A}{r^2\varSigma}W+\Psi(1-W), \label{Cin}
\end{eqnarray}
where $\Psi$ is a positive constant, whereas the conformal factor $\phi(r,\theta)$ will be determined 
by solving the constraint equations.  

Since the trace of the extrinsic curvature vanishes in $r> R$, we assume 
the same situation in $r<R$. Then, we write the extrinsic curvature in the form
\begin{equation}
K^{ij}=\phi^{-10}\left[X^{j|i}+X^{i|j}-\frac{2}{3}\lambda^{ij}X^k{}_{|k}\right]
=:\phi^{-10}(LX)^{ij},
\label{exc}
\end{equation}
where $_{|j}$ is the covariant derivative with respect to the conformal metric 
$\lambda_{ij}$ and $^{|j}=_{|i}\lambda^{ij}$. 
Substituting the above expression into the momentum constraint (\ref{MC}) 
with $J_i=0$, we have
\begin{equation}
D_j K^{ij}=\phi^{-10}(LX)^{ij}{}_{|j}=0.  \label{MC-1}
\end{equation}
Here, we assume 
\begin{equation}
X^i=\left(0,0,X^\varphi(r,\theta)\right).
\label{X}
\end{equation}
This assumption leads to the similar nontrivial components 
of the extrinsic curvature to those outside the shell,  
\begin{equation}
K^{r\varphi}=\frac{1}{\phi^{10}\A}\frac{\partial X^\varphi}{\partial r}~~~~~{\rm and}~~~~~
K^{\theta\varphi}=\frac{1}{r^2\phi^{10}\B}\frac{\partial X^\varphi}{\partial \theta}. \label{exc-in}
\end{equation}
Substituting Eq.~(\ref{exc-in}) into the momentum constraint (\ref{MC-1}), we have
\begin{equation}
\frac{\partial}{\partial r}\left(r^4\sqrt{\frac{\B\C^3}{\A}}\frac{\partial X^\varphi}{\partial r}\right)
+\frac{1}{\sin^3\theta}\frac{\partial}{\partial \theta }\left(r^2\sin^3\theta\sqrt{\frac{\A\C^3}{\B}}
\frac{\partial X^\varphi}{\partial \theta}\right)=0. 
\label{MC-2}
\end{equation}
The above equation is an elliptic type differential equation for $X^\varphi$. It is a practically very 
important fact that there is no conformal factor $\phi$ in Eq.~(\ref{MC-2}): We can solve Eq.~(\ref{MC-2})  
without solving the Hamiltonian constraint (\ref{HC}). 

In order to get a meaningful solution of Eq.~(\ref{MC-2}), we should impose 
an appropriate boundary condition on $r=R$. If we impose the 
continuity of the extrinsic curvature across $r=R$, we have from Eqs.~(\ref{exc-out}) and (\ref{exc-in}) 
the following two conditions:
\begin{eqnarray}
\frac{\partial X^\varphi}{\partial r}&=&\frac{1}{2\alpha}\frac{\partial \beta^\varphi}{\partial r}, \label{rp-comp}\\
\frac{\partial X^\varphi}{\partial \theta}&=&\frac{1}{2\alpha}\frac{\partial \beta^\varphi}{\partial \theta}.
\label{tp-comp}
\end{eqnarray}
The former condition comes from the continuity of $K^{r\varphi}$, whereas the latter one 
comes from $K^{\theta\varphi}$. Since the other components vanish identically in the both inside 
and outside of the shell, the continuities of those components are trivially guaranteed. 
The condition (\ref{rp-comp}) is the Neumann type, whereas the condition (\ref{tp-comp}) 
is the Dirichlet type since, by integrating (\ref{tp-comp}) with respect to $\theta$, we obtain 
\begin{equation}
X^\varphi(R,\theta)=\frac{a}{R^2+a^2}\left[\alpha(R,\theta)-\alpha(R,0)\right], \label{Xphi}
\end{equation}
where we have chosen the integration constant so that $X^\varphi(R,0)=0$. 
We cannot impose both of them at once; we will adopt 
the boundary condition (\ref{tp-comp}) or equivalently (\ref{Xphi}). 
In the next section, the reason why we adopt the condition (\ref{Xphi}) will be  made clear. 
Although our choice leads to the discontinuity of $K^{r\varphi}$ at $r=R$, no problem occurs. 
In accordance with Israel's formalism\cite{Israel}, the derivative normal to the shell  
of the spacetime metric does not have to be single valued 
on the shell but only be finite there. Since the extrinsic curvature $K_{ij}$ may contain 
the derivative normal to the shell, it can be discontinuous at $r=R$.

Substituting Eqs.~(\ref{metric}) and (\ref{exc}) into Eq.~(\ref{HC}) with $\rho=0$, 
we obtain an elliptic type differential equation for the conformal factor $\phi$: 
\begin{equation}
\phi^{|i}{}_{|i}-\frac{1}{8}{\cal R}\phi
+\frac{1}{8}\lambda_{il}\lambda_{jm}(LX)^{ij}
(LX)^{lm}\phi^{-7}=0,
\label{HC-1}
\end{equation}
where ${\cal R}$ is the Ricci scalar of the conformal metric $\lambda_{ij}$. 
After some manipulations, we obtain  
\begin{eqnarray}
&&\phi^{|i}{}_{|i}=\frac{1}{r^2\sqrt{\A\B\C}}\frac{\partial}{\partial r}
\left(r^2\sqrt{\frac{\B\C}{\A}}\frac{\partial\phi}{\partial r}\right)
+\frac{1}{r^2\sqrt{\A\B\C}\sin\theta}\frac{\partial}{\partial \theta}
\left(\sin\theta\sqrt{\frac{\A\C}{\B}}\frac{\partial\phi}{\partial \theta}\right), \\
&&\cr
&& {\cal R}=\frac{2}{r^2\B}
-\frac{1}{\A}\left[\frac{\partial^2_r (r^2\B)}{r^2\B}+\frac{\partial^2_r (r^2\C)}{r^2\C}\right] \cr
&&+\frac{1}{2\A}\Biggl[
\frac{\partial_r \A}{\A}\frac{\partial_r (r^2 \B)}{r^2\B}+\left(\frac{\partial_r (r^2 \B)}{r^2\B}\right)^2
+\frac{\partial_r \A}{\A}\frac{\partial_r(r^2 \C)}{r^2\C}
+\left(\frac{\partial_r(r^2 \C)}{r^2\C}\right)^2
-\frac{\partial_r (r^2\B)}{r^2\B}\frac{\partial_r (r^2 \C)}{r^2\C}
\Biggr] \cr
&&-\frac{1}{r^2\B}\left(\frac{\partial^2_\theta \A}{\A}+\frac{\partial^2_\theta \C}{\C}\right)
+\frac{1}{2\B}\Biggl[
\frac{\partial_\theta \A}{\A}\frac{\partial_\theta \B}{\B}+\left(\frac{\partial_\theta \A}{\A}\right)^2
+\frac{\partial_\theta \B}{\B}\frac{\partial_\theta \C}{\C}
+\left(\frac{\partial_\theta \C}{\C}\right)^2
-\frac{\partial_\theta \A}{\A}\frac{\partial_\theta \C}{\C}
\Biggr] \cr
&&-\frac{\cot\theta}{r^2\B}\left(
\frac{\partial_\theta \A}{\A}-\frac{\partial_\theta \B}{\B}
+2\frac{\partial_\theta \C}{\C}
\right), \\
&&\cr
&&\frac{1}{8}\lambda_{il}\lambda_{jm}(LX)^{ij}(LX)^{lm}
=\frac{1}{4}r^2\C\sin^2\theta\left[\frac{1}{\A}\left(\frac{\partial X^\varphi}{\partial r}\right)^2
+\frac{1}{r^2\B}\left(\frac{\partial X^\varphi}{\partial \theta}\right)^2\right].
\end{eqnarray}
Since the intrinsic metric $\gamma_{ij}=\phi^4h_{ij}$ 
should be continuous at $r=R$, the boundary condition 
on Eq.~(\ref{HC-1}) should be the following Dirichlet type:  
$$
\phi(R,\theta)=1.
$$ 

By solving Eqs.~(\ref{MC-2}) and (\ref{HC-1}), we can determine the initial values of 
$\gamma_{ij}$ and $K_{ij}$ in $r<R$. Here, we should note that the first order 
derivative of $\phi$ with respect to $r$ will be discontinuous at $r=R$. 
Since every nonvanishing component of 
the conformal metric $\lambda_{ij}$ is $C^5$ function [see Eqs.~(\ref{W-def})--(\ref{Cin})], the first order 
derivative of the intrinsic metric 
$\gamma_{ij}$ with respect to $r$ will be discontinuous at $r=R$. 

Since we do not impose the boundary condition (\ref{rp-comp}) on Eq.~(\ref{MC-2}), 
$K^{r\varphi}$ will not be single valued at $r=R$. 
The discontinuity of the first order derivative of $\gamma_{ij}$ 
and $K^{r\varphi}$ at $r=R$ 
implies the existence of a distributional source at $r=R$ in  
accordance with Israel's formalism\cite{Israel}. 
In the next section, we see what kinds of matter are confined on $r=R$.

\section{The surface stress-energy tensor}

\begin{figure}
\begin{center}
\includegraphics[width=0.5\textwidth]{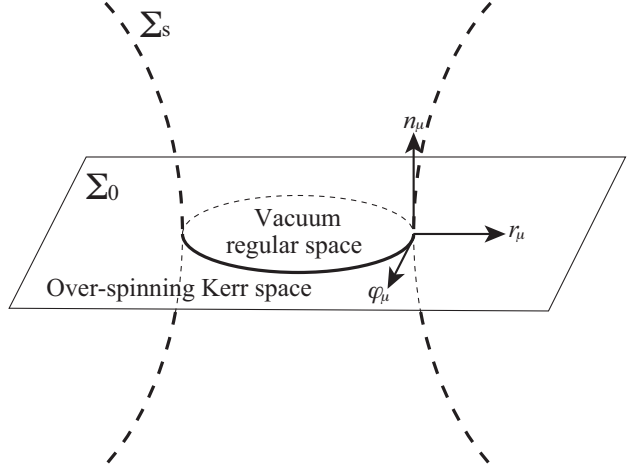}
\caption{\label{situation}
The situation we consider is depicted. Note that the trajectory of the infinitesimally thin shell depicted in this figure 
is speculative since we have not studied its dynamical evolution. Our interest  is 
focused on the only moment at which the size of the shell is extremum.   }
\end{center}
\end{figure}

The world volume of an infinitesimally thin shell will be a singular 
timelike hypersurface $\Sigma_{\rm s}$. 
We assume that the initial data corresponds to a moment at which the size of the shell 
is extremum: It may be a moment of a bounce due to its large angular momentum. 
This assumption implies that the timelike unit vector $n_\mu$ normal to 
the spacelike hypersurface 
$\Sigma_0$ is tangent to $\Sigma_{\rm s}$ (see Fig.~\ref{situation}). Here note that 
this assumption restricts not only the initial situation but also partly the time evolution. 
As a consequence of this assumption, the stress of the matter field confined in  
the shell is partly restricted 
although the information about the stress of the matter field is not necessary 
for setting up the initial data. 

The projection operator to $\Sigma_{\rm s}$ is defined as
\begin{equation}
h_\mu^\nu=\delta_\mu^\nu-r_\mu r^\nu,
\end{equation}
where $r_\mu$ is the unit vector normal to $\Sigma_{\rm s}$. 
The extrinsic curvature of $\Sigma_{\rm s}$ is defined as
\begin{equation}
Q_{\mu\nu}:=h_\mu^\rho h_\nu^\lambda \nabla_\rho r_\lambda.
\end{equation}
Then, Israel's condition of the metric junction is given by 
\begin{equation}
\left[Q_{\mu\nu}-h_{\mu\nu}Q^\rho_\rho\right]_\pm=-8\pi S_{\mu\nu},
\label{j-cond}
\end{equation}
where, denoting a quantity evaluated just outside the shell by a symbol with the subscript $+$ and 
that evaluated just inside the shell by the symbol with the subscript $-$, we have defined 
\begin{equation}
[\Psi]_\pm:=\Psi_+-\Psi_-.
\end{equation}
The quantity $S_{\mu\nu}$ on the right-hand side of Eq.~(\ref{j-cond}) 
can be regarded as the surface stress-energy tensor of the shell through 
Einstein's equations. 

We introduce a tetrad basis on the shell, which is composed of the unit vector 
$n_\mu$ normal to the spacelike hypersurface $\Sigma_0$ and  
\begin{eqnarray}
r_\mu&=&(0,\phi^2\sqrt{\A},0,0), \\
\theta_\mu&=&(0,0,r\phi^2\sqrt{\B},0), \label{theta-def}\\
\varphi_\mu&=&\left(0,0,0,r\phi^2\sqrt{\C}\sin\theta\right). \label{varphi-def}
\end{eqnarray}
By using this tetrad basis, the projection operator is written in the form 
\begin{equation}
h_{\mu\nu}=-n_\mu n_\nu+\theta_\mu\theta_\nu+\varphi_\mu\varphi_\nu.
\end{equation}

Through straightforward manipulations from Eq.~(\ref{j-cond}), 
we obtain the tetrad components of $S_{\mu\nu}$ as 
\begin{eqnarray}
S_{(n)(n)}
&=&-\frac{1}{8\pi}\left[Q_{(\theta)(\theta)}+Q_{(\varphi)(\varphi)}\right]_\pm \cr
&&\cr
&=&-\frac{1}{8\pi r^2\sqrt{\A\B\C}}\left[\frac{\partial (r^2\phi^4\sqrt{\B\C})}{\partial r} \right]_\pm 
=\frac{1}{2\pi\sqrt{\A}}~\frac{\partial \phi}{\partial r}\biggl|_{r=R-0}, \\
&&\cr
S_{(n)(\theta)}&=&-\frac{1}{8\pi}\left[Q_{(n)(\theta)}\right]_\pm
=-\frac{r\sqrt{\A\B}}{8\pi}\left[K^{r\theta}\right]_\pm=0, 
\label{nt-comp}\\
&&\cr
S_{(n)(\varphi)}&=&-\frac{1}{8\pi}\left[Q_{(n)(\theta)}\right]_\pm \cr
&&\cr
&=&-\frac{r\sqrt{\A\C}}{8\pi}\sin\theta\left[K^{r\varphi}\right]_\pm 
=-\frac{r}{8\pi}\sqrt{\frac{\C}{\A}}\sin\theta
\left(\frac{1}{2\alpha}\frac{\partial \beta^\varphi}{\partial r} 
-\frac{\partial X^\varphi}{\partial r}\right)\biggl|_{r=R}.
\label{Jvarphi}
\end{eqnarray}
Since we have
\begin{equation}
\left[Q_{(\theta)(\theta)}\right]_\pm
=\left[Q_{(\varphi)(\varphi)}\right]_\pm
=-\frac{2}{\sqrt{\A}}\frac{\partial\phi}{\partial r}\biggl|_{r=R-0},
\end{equation}
and
\begin{equation}
\left[Q_{(\theta)(\varphi)}\right]_\pm=0, 
\end{equation}
from Eq.~(\ref{j-cond}), we obtain
\begin{eqnarray}
S_{(\theta)(\theta)}
&=&-\frac{1}{8\pi}\left[Q_{(n)(n)}-Q_{(\varphi)(\varphi)}\right]_\pm \cr
&=&-\frac{1}{8\pi}\left[Q_{(n)(n)}-Q_{(\theta)(\theta)}\right]_\pm
=S_{(\varphi)(\varphi)}, \label{Stt}\\
S_{(\theta)(\varphi)}&=&0.
\end{eqnarray}
The above results imply that the surface stress-energy tensor takes the 
following form; 
\begin{equation}
S_{\alpha\beta}=\sigma n_\alpha n_\beta +j \left(n_\alpha \varphi_\beta+\varphi_\alpha n_\beta\right)
+p\left(\theta_\alpha \theta_\beta+\varphi_\alpha \varphi_\beta\right),
 \label{stress-energy}
\end{equation}
where
\begin{eqnarray}
\sigma
&:=&\frac{1}{2\pi\sqrt{\A}}~\frac{\partial \phi}{\partial r}\biggl|_{r=R-0}, \label{sigma-def}\\
&&\cr
j&:=&\frac{r}{8\pi}\sqrt{\frac{\C}{\A}}\sin\theta
\left(\frac{1}{2\alpha}\frac{\partial \beta^\varphi}{\partial r} 
-\frac{\partial X^\varphi}{\partial r}\right)\biggl|_{r=R}. \label{j-def}
\end{eqnarray}
It should be noted that $p$ has never been determined yet since $Q_{(n)(n)}$ 
is not restricted at all in the present initial data [see Eq.~(\ref{Stt})]. 

From the above results, we can see that the junction condition (\ref{j-cond}) 
does not impose the discontinuity of $K^{\theta\varphi}$ at $r=R$. This fact 
implies that $K^{\theta\varphi}$ should be everywhere continuous.  
Hence we should impose the Dirichlet boundary condition (\ref{tp-comp}) 
on the momentum constraint (\ref{MC-2}). 
From Eq.~(\ref{nt-comp}), we see that if $K^{r\varphi}$ is 
double valued at $r=R$, there is a nonvanishing 
angular momentum density $j$. 

The obtained solutions should satisfy the following conditions.
The conformal factor $\phi$ should be positive and finite in $r<R$. 
In Appendix A, we discuss 
the weak energy condition (WEC), the strong energy condition (SEC) 
and the dominant energy condition (DEC)~\cite{Wald} in the case of 
the present surface stress-energy tensor (\ref{stress-energy}). 
As shown in Appendix A, all of WEC, SEC and DEC are satisfied only if  
\begin{equation}
\sigma\geq \sqrt{2}|j|. \label{energy-c}
\end{equation}
As long as the above inequality holds, the appropriate $p$ guarantees all of the energy conditions.  
If the equality in Eq.~(\ref{energy-c}) is satisfied, $p$ should be equal to $\sigma/2$, and $S^{\alpha\beta}$ takes 
the following form;  
\begin{equation}
S^{\alpha\beta}=\frac{\sigma}{2}\left(v_+^\alpha v_+^\beta+\theta^\alpha \theta^\beta \right),
\end{equation}
where
\begin{equation}
v_+^\alpha=\sqrt{2}n^\alpha+\frac{j}{|j|}\varphi^\alpha.
\end{equation}
The detail of the derivation is shown in Appendix A. 

\section{Numerical result and Discussion}

\begin{figure}[h!]
\begin{center}
\includegraphics[width=0.5\textwidth]{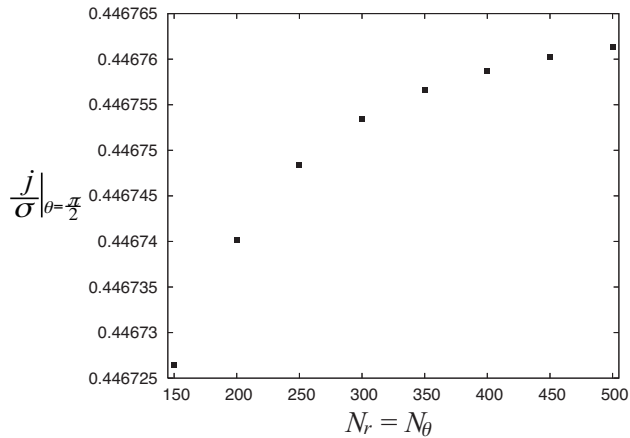}
\caption{\label{conv}
The value of $j/\sigma$ at $\theta=\pi/2$ 
is depicted as a function of the grid number $N_r=N_\theta$.    
The Kerr parameter $a$, the radius of the shell $R$ and the central value of the conformal factor 
$\Psi$ are, respectively, equal to $2M$, $0.75M$ and $0.2$. }
\end{center}
\end{figure}

We numerically solve the constraint equations (\ref{MC-2}) and (\ref{HC-1}). 
We solve the momentum constraint (\ref{MC-2}) first, and then, after   
substituting the solution of (\ref{MC-2}) into the Hamiltonian constraint (\ref{HC-1}), we solve 
Eq.~(\ref{HC-1}). 
We adopt the finite difference method of the second order accuracy.   
We denote the grid number for the domain $0<r<R$ by $N_r$  and for the domain 
$0< \theta <\pi/2$ by $N_\theta$. 
We take $N_r=1000$ and $N_\theta=100$ in typical run, but $N_r=2000$ and $N_\theta=200$ 
in the case that $\Psi$, $L/R$ or $R/M$ is small. The reason why the grid number in the 
$r$ direction is much larger than that for the $\theta$ direction is that the Ricci scalar ${\cal R}$ is a 
very steep function of $r$. 
We invoke the incomplete LU conjugate gradient squared method for 
the matrix inversions to solve elliptic type differential equations.   
In order to check the numerical code, we have seen the convergence of solutions with 
grid number increased:  See Fig. \ref{conv}. 

\begin{figure}
\begin{center}
\includegraphics[width=0.5\textwidth]{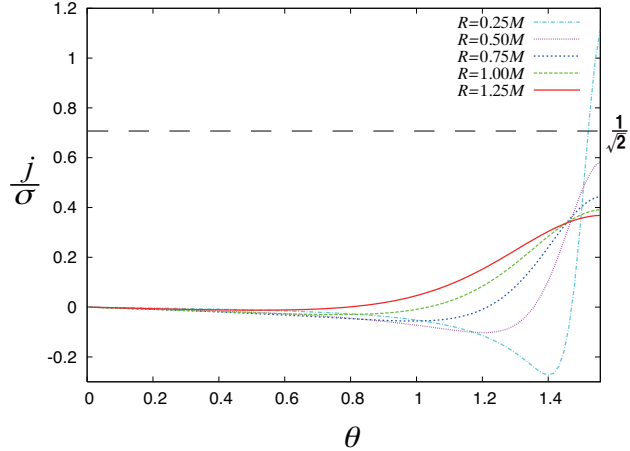}
\caption{\label{theta-ratio}
The ratio of $j$ to $\sigma$ is depicted as a function of $\theta$ for various $R$. The Kerr parameter $a$ and 
the parameter of the conformal metric $\Psi$ are, respectively, equal to $2M$ and $0.2$.  }
\end{center}
\end{figure}

The surface energy density $\sigma$ is positive in all our numerical calculations. 
In Fig.~\ref{theta-ratio}, the ratio of $j$ to $\sigma$ is depicted as a function of 
$\theta$ for various radii $R$ with 
$a=2M$, $L=0.5R$ and $\Psi=0.2$.  
We see from this figure that the maximal value of $|j|/\sigma$ is equal to the value of $j/\sigma$ 
at the equator $\theta=\pi/2$. This is true for all our numerical calculations. 

The other important tendency is that the smaller the radius of the shell $R$ is, 
the larger the maximal value of $|j|/\sigma$. We can see from Fig.~\ref{theta-ratio} that 
$\sigma$ is less than $\sqrt{2}|j|$ near the equator $\theta=\pi/2$ in 
the case of $R=0.25M$, 
whereas  the inequality (\ref{energy-c}) is satisfied for all $\theta$ in the cases of 
$R=0.5M$, $0.75M$, $M$  and $1.25M$. 
Hence, if we impose the reasonable energy conditions, i.e., Eq.~(\ref{energy-c}) with $L$ and $\Psi$ fixed, 
the radius $R$ is bounded below by some positive value. 
However as shown below, the lowest value of $R$ can be made smaller
by taking adequate value of $\Psi$ or $L$. 

\begin{figure}
\begin{center}
\includegraphics[width=0.5\textwidth]{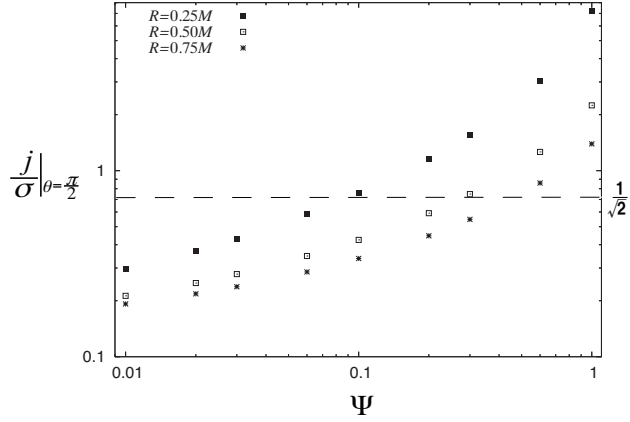}
\caption{\label{psi-ratio}
The value of $j/\sigma$ at $\theta=\pi/2$, which is equal to the maximal value of $|j|/\sigma$, is 
depicted as a function of the central value of the 
conformal factor $\Psi$ for three cases, $R=0.25M$, $0.5M$ and $0.75M$.   
The Kerr parameter $a$ is equal to $2M$, and $L=0.5R$ is assumed.}
\end{center}
\end{figure}
\begin{figure}
\begin{center}
\includegraphics[width=0.5\textwidth]{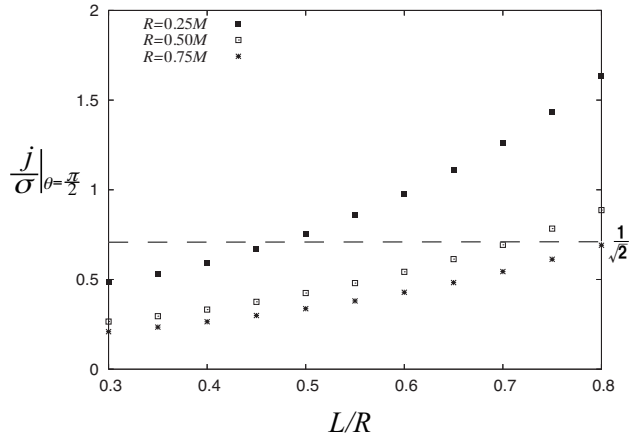}
\caption{\label{L-ratio}
The value of $j/\sigma$ at $\theta=\pi/2$, which is equal to the maximal value of $|j|/\sigma$, is 
depicted as a function of the smoothing length $L$ of the smoothed step 
function $W(r)$ for three cases, $R=0.25M$, $0.5M$ and $0.75M$.   
The Kerr parameter $a$ is equal to $2M$, and $\Psi=0.1$ is assumed.  }
\end{center}
\end{figure}

In Fig.~\ref{psi-ratio}, we 
depict the maximal value of $|j|/\sigma$, i.e., $j/\sigma$ at $\theta=\pi/2$ as a function 
of $\Psi$ for three cases $R=0.25M$, $0.5M$ and $0.75M$, where we assume 
$L=0.5R$ and $a=2M$. 
We see from this figure that the smaller $\Psi$ is, the smaller 
the maximal value of $|j|/\sigma$. It is worthwhile to notice that  
the energy condition (\ref{energy-c}) is satisfied even in the case of $R=0.25M$ 
if we choose $\Psi\lesssim9\times10^{-2}$.  

In Fig.~\ref{L-ratio}, we depict $j/\sigma$ at $\theta=\pi/2$ as a function of $L$ for three cases 
$R=0.25M$, $0.5M$ and $0.75M$, where we 
assume $\Psi=10^{-1}$. We can see from this figure that the smaller $L$ is, the 
smaller the maximal value of $|j|/\sigma$, i.e., $j/\sigma$ at $\theta=\pi/2$. 

In Appendices B and C, we discuss the behavior of the solutions of the momentum 
and Hamiltonian constraints in the limit of $L\rightarrow 0$ and show that $\sigma$ may become 
arbitrarily large in this limit, whereas  $|j|$ is bounded above. This means that 
the energy conditions may hold despite the value of  $R$, $a$ and $\Psi$ 
as long as $R>0$, $a>M$ and $0<\Psi<1$, if $L$ takes a sufficiently small value. 
Our numerical results are consistent with these estimates, but due to the limitation of the 
numerical resolution, we have not yet seen the asymptotic behavior expected from the discussions in Appendices B and C. Hence, exactly speaking, it is still an open question whether the energy conditions 
necessarily hold for sufficiently small $L$, but no lower bound on $R$ has been found 
in our numerical results. 

\begin{figure}
\begin{center}
\includegraphics[width=0.5\textwidth]{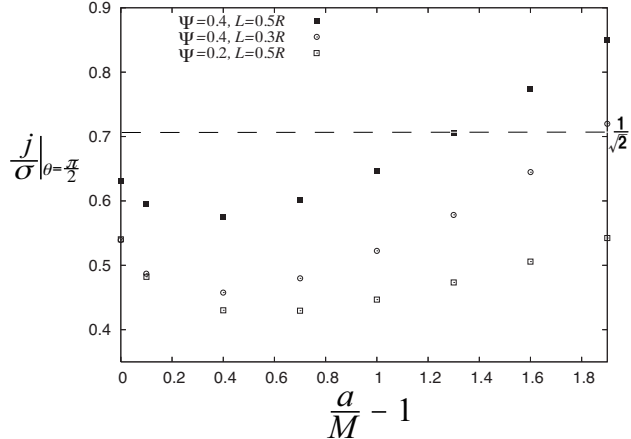}
\caption{\label{akerr-ratio}
The value of $j/\sigma$ at $\theta=\pi/2$ is 
depicted as a function of $a/M-1$; the smallest value of $a/M-1$ is zero. 
The radius of the shell $R$ is equal to $0.75M$. We depict three cases: 
the first one is of $\Psi=0.4$ with $L=0.5R$, 
the second one is of $\Psi=0.4$ with $L=0.3R$ and the third one is of $\Psi=0.2$ with $L=0.5R$. }
\end{center}
\end{figure}
Figure \ref{akerr-ratio} depicts $j/\sigma$ at $\theta=\pi/2$ 
as a function of $a/M-1$ with $R=0.75M$ in the three cases of $L=0.5R$ with $\Psi=0.2$, 
$L=0.5R$ with $\Psi=0.4$ and 
$L=0.3R$ with $\Psi=0.4$. 
We can see from this figure that $|j|/\sigma$ is decreasing for $a/M-1<0.4$ and increasing 
for $a/M-1>0.4$. It is the most important fact that Eq.~(\ref{energy-c}) is satisfied in the domain 
arbitrarily close to $a=M$ and even at $a=M$\footnote{In the case of $a=M$, the conformal metric $\lambda_{ij}$ 
is singular at $r=M$, since $\varDelta$ vanishes.  
However, in the case of $R<M$, since $\varDelta>0$ for $0\leq r\leq R$, 
the conformal metric $\lambda_{ij}$ is not singular inside and on the shell. 
Hence there is no singular point in the Hamiltonian and momentum constraints in the domain 
that we should know. }.

In the case of $a=M(1+\epsilon)$ with $0<\epsilon \ll1$, the value 
of $\varDelta(r)/M^2$ at $r=M$ 
is equal to $\epsilon(2+\epsilon)$ which is much less than unity. 
This implies that, as pointed out by Patil and Joshi\cite{PJ2011}, collisions of test particles 
with the trans-Planckian energy in their center of mass frame  
occur at $r\simeq M$ in the Kerr domain, since the collision energy defined in the 
center of mass frame of the particles is proportional to 
$\varDelta^{-1/2}$ at their collision event. 
Here we should note that, in contrast with 
the situation supposed by Patil and Joshi, the initial data 
we consider will not be a snapshot of a stationary configuration, and hence 
the trans-Planckian collisions of test particles may be allowed for only a short time interval 
in the present case.  

Finally, it is worthwhile to notice that 
the geometrical size of the shell is not necessarily small even in the case of $R\ll  M$.  
Even in the limit of $R\rightarrow0$, the geometrical size of the shell 
is finite: See Appendix D. Our result is consistent with the hoop conjecture which states that 
a black hole with horizon forms when and only when the mass $M$ gets compacted 
in the region whose circumference $C$ measured in every direction satisfies 
$C\leq 4\pi M$\cite{Hoop}.

\section{Summary and Discussion}

We studied how small a rapidly rotating body can be by investigating the initial data which is  
a snapshot of a rotating infinitesimally thin shell with a spherical topology: The exterior of the shell 
is set up so that its intrinsic and extrinsic geometries are completely the same as that of the Kerr 
spacetime with the over-threshold angular momentum $a>M$, whereas the interior of the shell 
is determined by solving numerically the constraint equations. 
In this set of initial data, the shell is located 
on $r=R$, where $r$ is the radial coordinate in the Boyer-Lindquist coordinate system and 
$R$ is a positive constant. 

In the present numerical results, no lower bound on $R$ of the over-spinning shell 
has been found. 
This result suggests that the cosmic censorship conjecture does not forbid  
the phenomenon similar to the Patil-Joshi process, i.e.,  the ultrahigh energy collision of particles 
due to the deep gravitational potential near the Kerr naked singularity. 

\begin{figure}[h!]
\begin{center}
\includegraphics[width=0.8\textwidth]{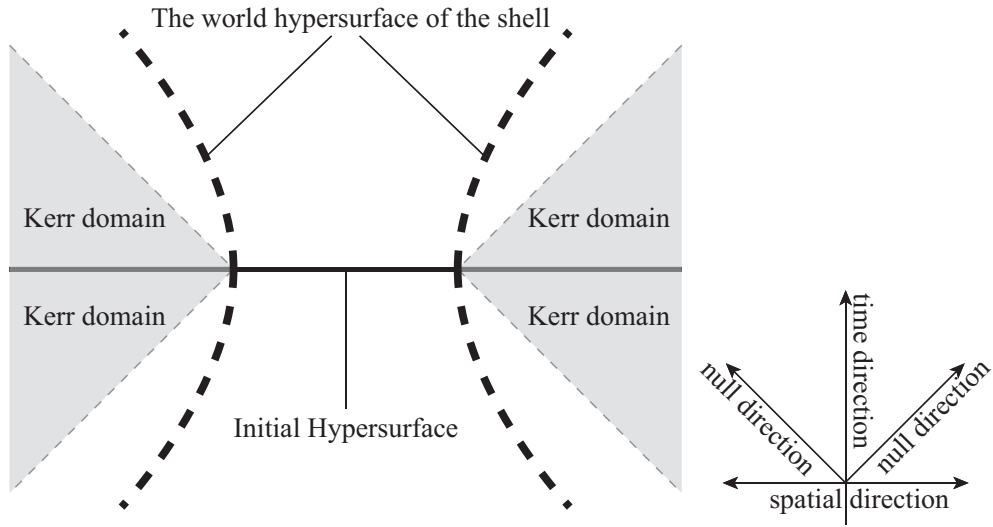}
\caption{\label{Kerr-domain}
The spacetime diagram of the present system. The timelike hypersurface of the shell denoted by dashed lines 
is almost speculative, since we have not solved the evolution of the shell. 
However the existence of the Kerr domains depicted in this figure by shaded regions
is not a speculation but a definite fact. 
 }
\end{center}
\end{figure}

Since the exterior domain of the shell is the same as a spacelike hypersurface 
of the over-spinning Kerr spacetime, the Kretschmann invariant $K$ 
in the exterior of the shell is given as
\begin{equation}
K\equiv R_{abcd}R^{abcd}=C_{abcd}C^{abcd}=
\frac{48M^{2}(r^{6}-15a^{2}r^{4}\cos^{2}\theta
+15a^{4}r^{2}\cos^{4}\theta-a^{6}\cos^{6}\theta)}
{(r^{2}+a^{2}\cos^{2}\theta)^{6}},
\end{equation}
in the Boyer-Lindquist coordinates. 
Just outside the shell, $K$ diverges in the limit of $R\rightarrow 0$ at the equator $\theta=\pi/2$:  
This corresponds to well-known ring singularity of the Kerr 
spacetime\footnote{Just on the shell, the Ricci part of $K$  
diverges due to the distributional material field. 
However, as we know, it is tractable singularity by Israel's prescription. 
By contrast, the divergence of the Weyl part 
of $K$ in the limit of $R\rightarrow0$ is too strong to treat the shell by Israel's formalism. 
The serious spacetime singularity appears on the shell in the limit of $R\rightarrow0$.}.
The complementary set of the causal future of the shell 
is equivalent to the over-spinning Kerr spacetime (see Fig.~\ref{Kerr-domain}), 
just outside of the shell is not enclosed by the 
event horizon however large the Weyl invariants is there. 
Our present results suggest the possibility of the 
formation of a spacetime border\cite{border} by an over-spinning body.

From a point of view of the observation of superstringy effects, 
Gimon and Ho\v rva have discussed a superspinar, which is a compact object 
with the angular momentum larger than 
the threshold value $J_{\rm max}$\cite{GH2007}. 
They do not expect the astrophysical formation of a superspinar and suggest 
the formation in the very early Universe. Our present result suggests a possibility 
of a superspinar through the shrinkage  
of a massive body, since the stringy effect will be important in the neighborhood of the 
over-spinning shell with very small $R$. 
However we should note that there are a few studies suggesting the instability of the 
superspinar model based on the over-spinning 
Kerr solution\cite{Dotti1,Dotti2,Pani_etal}. 
The superspinar may be a transient configuration or the geometry around the superspinar 
may be very different from the over-spinning Kerr spacetime, although  
this issue is outside the scope of the present paper.

\section*{Acknowledgments}
K.N. is thankful to H. Reall for giving a crucial idea in the present study. KN is also 
thankful to H. Yoshino for   pointing out the issue on the size of the shell. 
K.N. and M.K. are grateful to H. Ishihara and colleagues in the group of elementary 
particle physics and gravity at Osaka City University for useful discussions.  
K.N. was supported in part
by Japan Society for the Promotion of Science (JSPS) 
Grant-in-Aid for Scientifc Research (C) (No.~25400265). 
M.K was partially supported by  JSPS Fellows 
(No.~23$\cdot$2182) and grant for research abroad from JSPS. 
T.H. was partially supported by the Grant-in-Aid No.~26400282 for Scientific Research
Fund of the Ministry of Education, Culture, Sports, Science and Technology, Japan.

\appendix

\section{Energy conditions}

We can rewrite Eq.~(\ref{stress-energy}) in the following diagonalized form; 
\begin{equation}
S^{\alpha\beta}=\frac{\omega}{|\omega|}\left(\lambda_+v_+^\alpha v_+^\beta
-\lambda_-v_-^\alpha v_-^\beta\right)
+p\theta^\alpha \theta^\beta,
\end{equation}
where
\begin{equation}
\omega:=\sigma+p, \label{omega-def}
\end{equation}
and, defining the following quantity
\begin{equation}
\mu:=\sqrt{\omega^2-4j^2}\geq0, \label{mu-def}
\end{equation}
$\lambda_\pm$ and $v_\pm^\alpha$ are given by
\begin{equation}
\lambda_\pm=\frac{1}{2}(2\sigma-\omega\pm\mu), \label{lambda-def}
\end{equation}
and
\begin{equation}
v_\pm^\alpha=\frac{1}{\sqrt{2\mu}}\left(\sqrt{|\omega\pm\mu|}~n^\alpha
+\frac{\omega j}{|\omega j|}\sqrt{|\omega\mp\mu|}~\varphi^\alpha\right).
\end{equation}
Since it is believed that the stress-energy tensor of a physically reasonable material 
field except for a null fluid has real eigenvalues, we assume 
\begin{equation}
|\omega|\geq 2|j|. \label{omg-cons}
\end{equation}
We can easily see
\begin{equation}
g_{\alpha\beta}v_\pm^\alpha v_\pm^\beta=\mp \frac{\omega}{|\omega|}. \label{norm}
\end{equation}
Equation (\ref{norm}) implies that 
$\lambda_+$ corresponds to the energy density in the case of $\omega>0$, whereas 
$\lambda_-$ corresponds to the energy density in the case of $\omega<0$. 

We see what restrictions are imposed on $\sigma$, $j$ and $p$ by the weak, strong and dominant 
energy conditions (see e.g., Ref.~\cite{Wald} about the energy conditions). 

\subsection{Weak energy condition}

We consider the case of $\omega<0$ first. 
As mentioned, in this case, $\lambda_-$ is the energy density, 
and hence the weak energy condition (WEC) is equivalent to the following set of inequalities:  
\begin{eqnarray}
\lambda_-&\geq& 0, \label{WEC-m1} \\
\lambda_--\lambda_+&\geq&0, \label{WEC-m2} \\
\lambda_-+p&\geq&0. \label{WEC-m3}
\end{eqnarray}
From Eqs.~(\ref{omega-def}), (\ref{mu-def}), (\ref{lambda-def}) 
and the assumption of $\omega<0$, we have
\begin{equation}
\lambda_-+p=\frac{1}{2}\left(\omega-\mu\right)<0.
\end{equation}
The above result implies that the inequality $\omega<0$ contradicts WEC. 

Hereafter we assume $\omega\geq 0$.  Then, Eq.~(\ref{omg-cons}) becomes
\begin{equation}
\omega>2|j|. \label{real}
\end{equation}
Since  $\lambda_+$ corresponds to the energy density in the case of $\omega>0$, 
WEC is equivalent to the following set of inequalities: 
\begin{eqnarray}
\lambda_+&\geq& 0, \label{WEC-p1} \\
\lambda_+-\lambda_-&\geq&0, \label{WEC-p2} \\
\lambda_++p&\geq&0. \label{WEC-p3}
\end{eqnarray}
From Eqs.~(\ref{omega-def}), (\ref{mu-def}) and (\ref{lambda-def}), we have
\begin{equation}
\lambda_+-\lambda_-=\mu~~~~~{\rm and}~~~~~~\lambda_++p=\frac{1}{2}(\omega+\mu).
\end{equation}
Since both of $\omega$ and $\mu$ are positive, 
the conditions (\ref{WEC-p2}) and (\ref{WEC-p3}) are necessarily satisfied. 
Thus, the set of inequalities (\ref{real}) and (\ref{WEC-p1}) is equivalent to WEC, or equivalently,  
\begin{eqnarray}
\omega&\geq&2|j|, \cr
p&\leq&\frac{1}{2}(\omega+\mu). \nonumber
\end{eqnarray}

\subsection{Strong energy condition}

In the case of $\omega<0$, the strong energy condition (SEC) is equivalent to the set of inequalities 
(\ref{omg-cons}), (\ref{WEC-m2}), (\ref{WEC-m3}) and 
\begin{equation}
\lambda_--\lambda_++p\geq0. 
\end{equation}
As in the case of WEC, the condition (\ref{WEC-m3}) contradicts the assumption $\omega<0$, and hence 
$\omega\geq0$ should hold. 
Then, SEC is equivalent to the set of inequalities 
(\ref{real}), (\ref{WEC-p2}), (\ref{WEC-p3}) and 
\begin{equation}
\lambda_+-\lambda_-+p\geq0.  \label{SEC}
\end{equation}
We have
\begin{equation}
\lambda_+-\lambda_-+p=\mu+p,
\end{equation}
and hence $\mu+p\geq0$ should be satisfied so that the condition (\ref{SEC}) holds. Since 
Eqs.~(\ref{WEC-p2}) and (\ref{WEC-p3}) are trivially satisfied, 
SEC is equivalent to the following set of inequalities,
\begin{eqnarray}
\omega&\geq&2|j|, \cr
p&\geq&-\mu. \nonumber
\end{eqnarray}

\subsection{Dominant energy condition}

In the case of $\omega<0$, the dominant energy condition (DEC) is equivalent to 
\begin{eqnarray}
\lambda_-&\geq& |\lambda_+|, \label{DEC-m1} \\
\lambda_-&\geq& |p|. 
\end{eqnarray}
Equation (\ref{DEC-m1})  leads to $\lambda_-\geq0$ and hence by definition of $\lambda_\pm$, 
we have $\lambda_+=\lambda_-+\mu\geq0$. However, the inequality $\lambda_+>\lambda_->0$ contradicts 
Eq.~(\ref{DEC-m1}). 
Hence, $\omega\geq0$ should hold, and DEC is equivalent to $\omega>2|j|$ and 
\begin{eqnarray}
\lambda_+&\geq& |\lambda_-|, \label{DEC-1} \\
\lambda_+&\geq& |p|. \label{DEC-2}
\end{eqnarray}
It is not so difficult to see that the inequality (\ref{DEC-1}) is satisfied if and only if $p\leq\omega/2$. 
If $p$ is negative, Eq.~(\ref{DEC-2}) is satisfied by virtue of the non-negativity of $\omega$. 
For $p\geq0$, Eq.~(\ref{DEC-2}) leads to $p\leq(\omega+\mu)/4$. Because of the inequality 
$\omega\geq\mu$, if $p\leq(\omega+\mu)/4$ is satisfied, $p\leq\omega/2$ is also satisfied. 
Thus, DEC is equivalent to the following set of inequalities,
\begin{eqnarray}
\omega&\geq&2|j|, \cr
p&\leq&\frac{1}{4}(\omega+\mu). \nonumber
\end{eqnarray}

\subsection{Intersection of WEC, SEC and DEC}

We can see that all of WEC, SEC and DEC are satisfied at once 
if and only if the following set of inequalities holds: 
\begin{eqnarray}
\omega&\geq&2|j|, \label{e-cond-1}\\
-\mu&\leq&p\leq\frac{1}{4}(\omega+\mu). \label{e-cond-2}
\end{eqnarray}

Equation (\ref{e-cond-1}) leads to
\begin{equation}
\sigma+p\geq2|j|. \label{e-cond-1-2}
\end{equation}

The half of Eq.~(\ref{e-cond-2}), i.e., $-\mu \leq p$ leads to $p\geq0$, or if $p<0$, then
\begin{equation}
p\geq -\frac{1}{2}\sigma+\frac{2j^2}{\sigma}. \label{e-cond-2-2}
\end{equation}
Hence, the intersection of Eq.~(\ref{e-cond-1-2}) and $-\mu<p$ is given by 
\begin{eqnarray}
\sigma\geq-p+\sqrt{p^2+4j^2}~~~~~&{\rm for}&~~p<0, \label{e-cond-2-3}\\
\sigma\geq-p+2|j|~~~~~~~~~~~~~~&{\rm for}&~~p\geq0. \label{e-cond-1-3}
\end{eqnarray}

The other half of Eq.~(\ref{e-cond-2}), i.e., $p\leq(\omega+\mu)/4$ leads to 
$\sigma > 3p$, or if $\sigma \leq 3p$, then since $\omega \leq 4p$, $p$ is necessarily 
non-negative, and hence we have
\begin{equation}
p\geq0~~~{\rm and}~~~\sigma\geq p+\frac{j^2}{2p}. \label{e-cond-2-5}
\end{equation}
Hence the intersection of Eqs.~(\ref{e-cond-1-2}) and $p\leq(\omega+\mu)/4$ is given by
\begin{eqnarray}
\sigma\geq-p+2|j|~~~~~~~&{\rm for}&~~p<\frac{|j|}{2}, \label{e-cond-1-4}\\
\sigma\geq p+\frac{j^2}{2p}~~~~~~~~~~~&{\rm for}&~~p\geq \frac{|j|}{2}. \label{e-cond-2-6}
\end{eqnarray}

As a result, the WEC, SEC and DEC are satisfied, if and only if the following set of inequalities 
is satisfied;
\begin{eqnarray}
\sigma\geq-p+\sqrt{p^2+4j^2}~~~~~&{\rm for}&~~p<0, \label{e-cond-2-7}\\
\sigma\geq-p+2|j|~~~~~~~~~~~~~~&{\rm for}&~~0\leq p <\frac{|j|}{2}, \label{e-cond-1-5}\\
\sigma\geq p+\frac{j^2}{2p}~~~~~~~~~~~~~~~~~&{\rm for}&~~p\geq \frac{|j|}{2}. \label{e-cond-2-8}
\end{eqnarray} 
We find that the minimal value of $\sigma$ is given by the positive minimum of the function 
$f(p)=p+j^2/2p$, i.e., $\sigma=f(|j|/\sqrt{2})=\sqrt{2}|j|$: all of WEC, SEC and DEC is satisfied 
by the appropriate choice of $p$, only if 
\begin{equation}
\sigma\geq\sqrt{2}|j|
\end{equation}
holds. 

\subsection{Positivity of the stress}

As mentioned, if we assume $\omega>0$, $-\lambda_-$ is the stress. We show the condition 
that the stress $-\lambda_-$ is non-negative in the domain specified by 
Eqs.~(\ref{e-cond-2-7})--(\ref{e-cond-2-8}).  
The condition of the non-negative $-\lambda_-$ is equivalent to 
the inequality, $\sigma-p\leq\mu$. In the intersection of $\sigma-p\leq\mu$ and the domain specified by 
Eqs. (\ref{e-cond-2-7})--(\ref{e-cond-2-8}), $\sigma-p$ is non-negative and hence we have 
$(\sigma-p)^2\leq\mu^2$, or equivalently, $p\sigma\geq j^2$. The intersection between $p\sigma\geq j^2$ and 
Eqs. (\ref{e-cond-2-7})--(\ref{e-cond-2-8}) is given by
\begin{eqnarray}
\sigma\geq \frac{j^2}{p}~~~~~~~~~~~~~~~~~~~~&{\rm for}&~~0<p\leq\frac{|j|}{\sqrt{2}},  \label{e^cond-3} \\
\sigma\geq p+\frac{j^2}{2p}~~~~~~~~~~~~~~&{\rm for}&~~p\geq \frac{|j|}{\sqrt{2}}, \label{e-cond-2-9}
\end{eqnarray}
and there is no intersection in the domain of $p\leq0$. 
In the domain specified by Eqs.~(\ref{e^cond-3}) and (\ref{e-cond-2-9}), all of WEC, SEC and DEC hold and 
both of the stresses $-\lambda_-$ and $p$ are positive. 

\section{Momentum constraint of $L\rightarrow0$}

By integrating Eq.~(\ref{MC-2}) from $r=R-L$ to $r=R$, we have
\begin{eqnarray}
\int_{R-L}^Rdr\frac{\partial}{\partial r}
\left(r^4\sqrt{\frac{\B\C^3}{\A}}\frac{\partial X^\varphi}{\partial r}\right)
&=& R^4\sqrt{\frac{A^3(R,\theta)\varDelta(R)}{\varSigma(R,\theta)}}
\frac{\partial X^\varphi}{\partial r}\biggr|_{r=R}
-(R-L)^4\sqrt{\Psi^3}
\frac{\partial X^\varphi}{\partial r}\biggr|_{r=R-L}
\cr
&&\cr
&=&-\int_{R-L}^R dr\frac{1}{\sin^3\theta}\frac{\partial}{\partial \theta }\left(r^2\sin^3\theta\sqrt{\frac{\A\C^3}{\B}}
\frac{\partial X^\varphi}{\partial \theta}\right). 
\label{delX-1}
\end{eqnarray}

In the limit of $L \rightarrow0$, the momentum constraint (\ref{MC-2}) takes a very simple form 
in the domain $0\leq r<R$:  
\begin{equation}
\frac{\partial}{\partial r}\left(r^4\frac{\partial X^\varphi}{\partial r}\right)
+\frac{1}{\sin^3\theta}\frac{\partial}{\partial \theta }\left(r^2\sin^3\theta
\frac{\partial X^\varphi}{\partial \theta}\right)=0. 
\label{MC-3}
\end{equation}
If we solve Eq.~(\ref{MC-3}) by assuming $X^\varphi|_{r=R-0}=X^\varphi|_{r=R}$ and 
imposing the boundary condition (\ref{Xphi}), 
we will get a regular solution in the domain $0\leq r \leq R$ with finite $\partial X^\varphi/\partial r|_{r=R-0}$. 
Furthermore, since even in the limit of $L\rightarrow0$, all metric variables and $X^\varphi$ and their derivatives 
with respect to $\theta$ will be finite in the domain $R-L <r<R$, the integral in the last equality 
of Eq.~(\ref{delX-1}) vanishes in the limit of $L\rightarrow0$. Thus, we have
\begin{equation}
\frac{\partial X^\varphi}{\partial r}\biggr|_{r=R}
=R^4\sqrt{\frac{\Psi^3 \varSigma^3(R,\theta)}{A^3(R,\theta)\varDelta(R)}}
\frac{\partial X^\varphi}{\partial r}\biggr|_{r=R-0}.
\label{delX-2}
\end{equation}
Equation~(\ref{delX-2}) implies that $\partial X^\varphi/\partial r|_{r=R}$ is finite, 
and hence the surface angular momentum density $j$ is finite even 
in the limit of $L\rightarrow0$: See Eq.~(\ref{j-def}).

Since $\partial X^\varphi/\partial r$ is finite in the neighborhood of $r=R$, we have
\begin{equation}
\lim_{L\rightarrow0}\int_{R-L}^R\frac{\partial X^\varphi}{\partial r} dr=X^\varphi|_{r=R}-X^\varphi|_{r=R-0}=0.
\end{equation}
The above result is consistent with our assumption $X^\varphi|_{r=R-0}=X^\varphi|_{r=R}$.  

\section{Hamiltonian constraint of $L\rightarrow0$}

We consider the behavior of the solution of the Hamiltonian constraint in the 
limit of $L\rightarrow0$. In this limit, the conformal metric $\lambda_{ij}$ 
is still everywhere finite and a smooth function of $\theta$ but discontinuous in the $r$ direction at $r=R$. 
We assume that the conformal factor $\phi$ is also everywhere finite and smooth with respect to $\theta$ 
but may be discontinuous along the $r$ direction at $r=R$ in the limit of $L\rightarrow0$. The consistency of this 
assumption will be considered later. 

By integrating Eq.~(\ref{HC-1}) from $r=R-L$ to $r=R$, we have
\begin{eqnarray}
\int_{R-L}^R dr\frac{\partial}{\partial r}\left(r^2\sqrt{\frac{\B\C}{\A}}\frac{\partial\phi}{\partial r}\right)
&=&\sqrt{\frac{A(R,\theta)\varDelta(R)}{\varSigma}}\frac{\partial\phi}{\partial r}\biggl|_{r=R}
-(R-L)^2\sqrt{\Psi}\frac{\partial\phi}{\partial r}\biggl|_{r=R-L} \cr
&&\cr
&=&T_1+T_2+T_3, \label{del-phi-1}
\end{eqnarray}
where
\begin{eqnarray}
T_1&=&\frac{1}{8}\int_{R-L}^Rdrr^2\sqrt{{\cal ABC}}{\cal R}\phi, \\
T_2&=&\frac{1}{8}\int_{R-L}^Rdrr^2\sqrt{{\cal ABC}}\lambda_{il}\lambda_{jm}(LX)^{ij}(LX)^{lm}\phi^{-7}, \\
T_3&=&-\int_{R-L}^Rdr\frac{1}{\sin\theta}\frac{\partial}{\partial \theta}
\left(\sin\theta\sqrt{\frac{\A\C}{\B}}\frac{\partial\phi}{\partial \theta}\right).
\end{eqnarray}
Integrals $T_2$ and $T_3$ vanish in the limit of $L \rightarrow 0$, since their integrands are finite, 
but $T_1$ does not, as shown below. We have 
\begin{eqnarray}
T_1&=&-\frac{1}{8}\int_{R-L}^Rdr\phi
\left[\sqrt{\frac{\C}{\A\B}}\partial^2_r (r^2\B)
+\sqrt{\frac{\B}{\A\C}}\partial^2_r (r^2\C)\right] \cr
&&\cr
&+&\frac{1}{16}\int_{R-L}^Rdrr^2\sqrt{\frac{\B\C}{\A}}\phi\Biggl[
\frac{\partial_r \A}{\A}\frac{\partial_r (r^2 \B)}{r^2\B}+\left(\frac{\partial_r (r^2 \B)}{r^2\B}\right)^2
+\frac{\partial_r \A}{\A}\frac{\partial_r(r^2 \C)}{r^2\C} \cr
&&\cr
&+&\left(\frac{\partial_r(r^2 \C)}{r^2\C}\right)^2
-\frac{\partial_r (r^2\B)}{r^2\B}\frac{\partial_r (r^2 \C)}{r^2\C}
\Biggr] \cr
&&\cr
&+&T_{12}, \label{T1-1}
\end{eqnarray}
where
\begin{eqnarray}
T_{12}&=&\frac{1}{4}\int_{R-L}^Rdr\phi \sqrt{\frac{\A\C}{\B}} 
-\frac{1}{8}\int_{R-L}^Rdr\phi\sqrt{\frac{\A\C}{\B}}
\left(\frac{\partial^2_\theta \A}{\A}+\frac{\partial^2_\theta \C}{\C}\right) \cr
&&\cr
&+&\frac{1}{16}\int_{R-L}^Rdrr^2\phi\sqrt{\frac{\A\C}{\B}}\Biggl[
\frac{\partial_\theta \A}{\A}\frac{\partial_\theta \B}{\B}+\left(\frac{\partial_\theta \A}{\A}\right)^2
+\frac{\partial_\theta \B}{\B}\frac{\partial_\theta \C}{\C}
+\left(\frac{\partial_\theta \C}{\C}\right)^2
-\frac{\partial_\theta \A}{\A}\frac{\partial_\theta \C}{\C}
\Biggr] \cr
&&\cr
&-&\frac{1}{8}\int_{R-L}^Rdr \phi \cot\theta\sqrt{\frac{\A\C}{\B}}
\left(
\frac{\partial_\theta \A}{\A}-\frac{\partial_\theta \B}{\B}
+2\frac{\partial_\theta \C}{\C}
\right).
\end{eqnarray}
By the assumption of the finiteness of $\phi$, $T_{12}$ vanishes in the limit of $L\rightarrow0$. 

By two times of the integration by part in Eq.~(\ref{T1-1}), we have
\begin{eqnarray}
T_1&=&-\frac{1}{8}
\left[r^2\phi\sqrt{\frac{\B\C}{\A}}\partial_r\ln(r^4\B\C)\right]_{r=R-L}^{r=R}
+\frac{1}{8}\int_{R-L}^Rdrr^2\sqrt{\frac{\B\C}{\A}}(\partial_r\phi)\partial_r\ln(r^4\B\C) \cr
&&\cr
&+&\frac{1}{16}\int_{R-L}^Rdrr^2\phi\sqrt{\frac{\B\C}{\A}}\partial\ln(r^2\B)\partial_r\ln(r^2\C) 
+T_{12}\cr
&&\cr
&=&-\frac{1}{8}
\left[r^2\phi\sqrt{\frac{\B\C}{\A}}\partial_r\ln(r^4\B\C)
-r^2\sqrt{\frac{\B\C}{\A}}(\partial_r\phi)
\ln\left(\frac{r^4\B(r,\theta)\C(r,\theta)}{R^4 \B(R,\theta)\C(R,\theta)}\right)\right]^{r=R}_{r=R-L} \cr
&&\cr
&-&\frac{1}{8}\int_{R-L}^Rdr\partial_r\left(r^2\sqrt{\frac{\B\C}{\A}}\partial_r\phi\right)
\ln\left(\frac{r^4\B(r,\theta)\C(r,\theta)}{R^4 \B(R,\theta)\C(R,\theta)}\right) \cr
&&\cr
&+&\frac{1}{16}\int_{R-L}^Rdrr^2\phi\sqrt{\frac{\B\C}{\A}}\partial\ln(r^2\B)\partial_r\ln(r^2\C) +T_{12}\cr
&&\cr
&=&-\frac{1}{8}
\left[r^2\phi\sqrt{\frac{\B\C}{\A}}\partial_r\ln(r^4\B\C)
-r^2\sqrt{\frac{\B\C}{\A}}(\partial_r\phi)
\ln\left(\frac{r^4\B(r,\theta)\C(r,\theta)}{A(R,\theta)}\right)\right]^{r=R}_{r=R-L} \cr
&&\cr
&-&\frac{1}{8}\int_{R-L}^Rdr\left[
-\frac{1}{\sin\theta}\partial_\theta\left(\sin\theta\sqrt{\frac{\A\C}{\B}}\partial_\theta\phi\right)
+\frac{1}{8}r^2\sqrt{\A\B\C}\left\{{\cal R}\phi-\lambda_{il}\lambda_{jm}(LX)^{ij}(LX)^{lm}\phi^{-7}\right\}
\right] \cr
&&\cr
&\times&\ln\left(\frac{r^4\B(r,\theta)\C(r,\theta)}{A(R,\theta)}\right) \cr
&&\cr
&+&\frac{1}{16}\int_{R-L}^Rdrr^2\phi\sqrt{\frac{\B\C}{\A}}\partial\ln(r^2\B)\partial_r\ln(r^2\C)+T_{12},
\label{T1-2}
\end{eqnarray}
where in the last inequality, we have used the Hamiltonian constraint (\ref{HC-1}). 
Then, from Eq.~(\ref{T1-2}), we have
\begin{eqnarray}
\int_{R-L}^R&drr^2&\sqrt{\A\B\C}\phi{\cal R}\left[
1+\frac{1}{8}\ln\left(\frac{r^4\B(r,\theta)\C(r,\theta)}{A(R,\theta)}\right)
\right] \cr
&&\cr
&=&-\sqrt{\frac{A\varDelta}{\varSigma}}\partial_rA\biggr|_{r=R}
+4(R-L)\sqrt{\Psi} \phi|_{r=R-L}
-(R-L)^2\sqrt{\Psi}\ln\left(\frac{(R-L)^4\Psi^2}{A(R,\theta)}\right)(\partial_r\phi)\biggr|_{r=R-L} \cr
&&\cr
&&-\int_{R-L}^Rdr\left[
-\frac{1}{\sin\theta}\partial_\theta\left(\sin\theta\sqrt{\frac{\A\C}{\B}}\partial_\theta\phi\right)
+\frac{1}{8}r^2\sqrt{\A\B\C}\lambda_{il}\lambda_{jm}(LX)^{ij}(LX)^{lm}\phi^{-7}
\right] \cr
&&\cr
&&\times\ln\left(\frac{r^4\B(r,\theta)\C(r,\theta)}{A(R,\theta)}\right) \cr
&&\cr
&&+\frac{1}{2}\int_{R-L}^Rdrr^2\phi\sqrt{\frac{\B\C}{\A}}\partial_r\ln(r^2\B)\partial_r\ln(r^2\C).
\label{T1-3}
\end{eqnarray}

Here it should be noted that, in the limit of $L \rightarrow0$, we have
\begin{eqnarray}
&&\partial_r\ln(r^2\B)=\frac{2r[(1-\Psi)W+\Psi]+(\varSigma-r^2\Psi)dW/dr}{(\varSigma-r^2\Psi)W+r^2\Psi} \cr
&&\cr
&&~~~~~\longrightarrow\frac{2r(1-\Psi)\vartheta(r-R)+2r\Psi}{(\varSigma-r^2\Psi)\vartheta(r-R)+r^2\Psi}
+\frac{2(\varSigma-R^2\Psi)}{\varSigma+R^2\Psi}\biggr|_{r=R}\delta(r-R), \label{drlnB} \\
&&\cr
&&\partial_r\ln(r^2\C)=\frac{[\partial_r(A/\varSigma)-2r\Psi]W+2r\Psi+(A/\varSigma-r^2\Psi)dW/dr}
{(A/\varSigma-r^2\Psi)W+r^2\Psi} \cr
&&\cr
&&~~~~~\longrightarrow\frac{[\partial_r(A/\varSigma)-2r\Psi]\vartheta(r-R)+2r\Psi}{(A/\varSigma-r^2\Psi)\vartheta(r-R)+r^2\Psi}
+\frac{2(A/\varSigma-R^2\Psi)}{A/\varSigma+R^2\Psi}\biggr|_{r=R}\delta(r-R), \label{drlnC}
\end{eqnarray}
where $\vartheta(x)$ is Heaviside's step function with $\vartheta(0)=1/2$, and $\delta(x)$ is Dirac's delta function. 
Hereafter, we choose $0<\Psi <1$. By this choice, the coefficients of Dirac's delta functions in 
Eqs.~(\ref{drlnB}) and (\ref{drlnC}) are always positive.  Then, we obtain
\begin{equation}
\lim_{L\rightarrow0}
\int_{R-L}^Rdrr^2\phi\sqrt{\frac{\B\C}{\A}}\partial_r\ln(r^2\B)\partial_r\ln(r^2\C)=+\infty. 
\label{div-integral}
\end{equation}
We should also note that, for $L\rightarrow0$,  
\begin{eqnarray}
\ln\left(\frac{r^4\B(r,\theta)\C(r,\theta)}{A(R,\theta)}\right)
\longrightarrow \ln \left(\frac{[A(r,\theta)-r^2\Psi]\vartheta(r-R)+r^4\Psi^2}{A(R,\theta)}\right).
\end{eqnarray}
Hence, we have, for $L\rightarrow0$, 
\begin{eqnarray}
&&\lim_{L\rightarrow0}\int_{R-L}^R drr^2\sqrt{\A\B\C}\phi{\cal R}\left[
1+\frac{1}{8}\ln\left(\frac{r^4\B(r,\theta)\C(r,\theta)}{A(R,\theta)}\right)
\right]  \cr
&&\cr
&&~~~~=\left[
1+\frac{1}{8}\ln\left(\frac{A(R,\theta)+R^4\Psi^2}{2A(R,\theta)}\right)
\right] 
\lim_{L\rightarrow0}\int_{R-L}^Rdrr^2\sqrt{\A\B\C}\phi{\cal R} 
\label{R-int}
\end{eqnarray}
We should note
\begin{equation}
1+\frac{1}{8}\ln\left(\frac{A(R,\theta)+R^4\Psi^2}{2A(R,\theta)}\right)
>1-\frac{1}{8}\ln2 >0.
\end{equation}

By the assumption of the smoothness of $\phi$ with respect to $\theta$, we have
\begin{equation}
\lim_{L\rightarrow0}\phi|_{r=R-L}=\Phi(\theta), \label{assmp}
\end{equation}
where $\Phi(\theta)$ is a smooth function of $\theta$. 
In the limit of $L\rightarrow0$, the Hamiltonian constraint in the domain $0 \leq r<R$ becomes
\begin{equation}
\frac{\partial}{\partial r}\left(r^2\frac{\partial\phi}{\partial r}\right)
+\frac{1}{\sin\theta}\frac{\partial}{\partial\theta}\left(\sin\theta\frac{\partial\phi}{\partial\theta}\right)=
\frac{1}{8}\Psi r^2\lambda_{il}\lambda_{jm}(LX)^{ij}(LX)^{lm}\phi^{-7}. \label{HC-2}
\end{equation}
Here we should note that $\lambda_{il}\lambda_{jm}(LX)^{ij}(LX)^{lm}$ is finite even in the limit 
of $L\rightarrow0$ (see Appendix B). 
Since Eq.~(\ref{assmp}) gives a Dirichlet boundary condition, the solution of Eq.~(\ref{HC-2}) 
with the finite radial derivative $\partial_r\phi|_{r=R-0}$ will exist. 
Hence, from Eqs.~(\ref{T1-2}), (\ref{div-integral}) and (\ref{R-int}), we have, for $0<L\ll R$,
\begin{eqnarray}
T_1 \simeq F_1(\theta)\int_{R-L}^Rdrr^2\phi\sqrt{\frac{\B\C}{\A}}\partial_r\ln(r^2\B)\partial_r\ln(r^2\C), \label{T1-3}
\end{eqnarray}
where $F_1(\theta)$ is a positive function of $\theta$. 
Equation (\ref{T1-3}) suggests 
\begin{equation}
\lim_{L\rightarrow0}T_1 =+\infty.
\end{equation}
Hence we see from Eq.~(\ref{del-phi-1}) that $\partial_r\phi|_{r=R}$ positively diverges for $L\rightarrow0$, 
but, for consistency of the assumption (\ref{assmp}), the following relation should hold: 
\begin{equation}
\lim_{L\rightarrow 0+}\int_{R-L}^R\frac{\partial\phi}{\partial r}dr=1-\Phi(\theta).
\label{consis}
\end{equation}

Here we will not present a rigorous proof of the consistency of the assumption 
imposed on $\phi$ at the beginning of this Appendix. However, 
on the ground of the dimensional analysis, this assumption seems to be reasonable. In the situation of 
$0<L \ll R$, we will have
\begin{equation}
\int_{R-L}^Rdrr^2\phi\sqrt{\frac{\B\C}{\A}}\partial_r\ln(r^2\B)\partial_r\ln(r^2\C)=\frac{F_2(\theta)}{L}
+{\cal O}(L^0),
\end{equation}
where $F_2(\theta)$ is a positive function of $\theta$. 
From Eqs.~(\ref{del-phi-1}), (\ref{T1-3}) and the above estimate, 
we have, for $0<L \ll R$, 
\begin{equation}
\int_{R-L}^R\frac{\partial\phi}{\partial r}dr=F_3(\theta)+{\cal O}(L),
\end{equation}
where $F_3$ is a function of $\theta$. 
Hence, the consistency condition (\ref{consis}) may hold. 

From the above considerations, we may have 
\begin{equation}
\frac{j}{\sigma} \longrightarrow 0~~~~~{\rm for}~~L \longrightarrow 0,
\end{equation}
since $\sigma$ is proportional to $\partial_r\phi|_{r=R}$ from Eq.~(\ref{sigma-def}). Hence, if we 
adopt sufficiently small $L$, the energy conditions may be satisfied despite the values of $R$ and $a$ as long as 
$R>0$ and $a>M$. 

\section{The geometrical size of the shell}

\begin{figure}
\begin{center}
\includegraphics[width=0.5\textwidth]{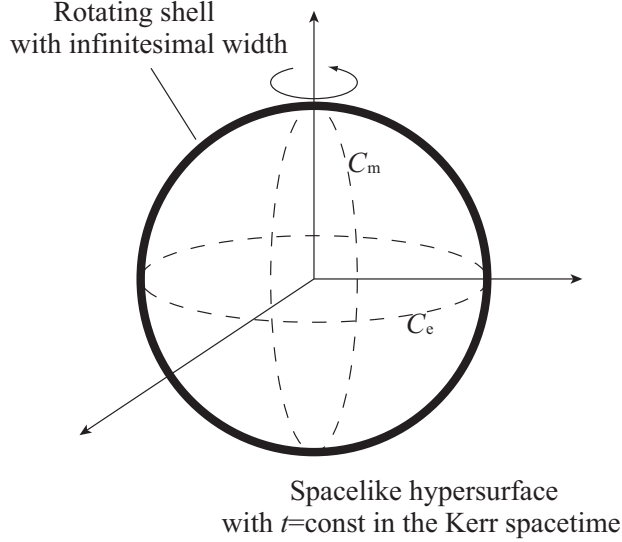}
\caption{\label{scale}
The circumferential $C_{\rm e}$ and $C_{\rm m}$ of the shell are depicted.  }
\end{center}
\end{figure}

The circumference $C_{\rm e}$ of the equator $\theta=\pi/2$ of the shell is given by
\begin{equation}
C_{\rm e}(R)=2\pi R\sqrt{{\cal C}(R,\pi/2)}=2\pi\sqrt{R^2+a^2+\frac{2Ma^2}{R}}.
\end{equation}
We can easily see that $C_{\rm e}$ takes a minimum value 
$2\pi\sqrt{3(Ma^2)^{2/3}+a^2}$ at $R=(Ma^2)^{1/3}$. If $a>M$, $C_{\rm e}$ is larger 
than $4\pi M$; this is consistent to 
the hoop conjecture\cite{Hoop}. The circumferential length in the meridian direction $C_{\rm m}$ is given by
\begin{equation}
C_{\rm m}(R)=4\int_0^{\pi/2} R\sqrt{{\cal B}(R,\theta)}d\theta =4\sqrt{R^2+a^2}
E\left(\sqrt{\frac{a^2}{R^2+a^2}}\right),
\end{equation}
where $E(k)$ is the complete elliptic integral of the second kind. Since $E(k)$ is 
monotonically decreasing with respect to $k$, the minimal value of $C_{\rm m}$ 
is equal to $4a$ achieved at $R=0$. The area of the shell $A_{\rm s}$ is given by
\begin{eqnarray}
A_{\rm s}&=&4\pi\int_0^{\pi/2}R^2\sqrt{{\cal B}(R,\theta) {\cal C}(R,\theta)}\sin\theta d\theta \cr
&&\cr
&=&2\pi\left[
R^2+a^2+\frac{R(R^3+a^2R+2Ma^2)}{2a\sqrt{\varDelta(R)}}
\ln\left|\frac{R^2+a^2+a\sqrt{\varDelta(R)}}{R^2+a^2-a\sqrt{\varDelta(R)}}\right|
\right].
\end{eqnarray}
The minimal value of $A_{\rm s}$ is equal to $2\pi a^2$ achieved at $R=0$. 
Hence, in this sense, the size of the shell is bounded below in the present case: see Fig \ref{scale}. 


\end{document}